\documentclass[twocolumn,showpacs,preprintnumbers,amsmath,amssymb]{revtex4}
\usepackage{graphicx}

\begin{document}
\def \ee {\varepsilon}
\thispagestyle{empty}
\title{Comment on ``Lifshitz-Matsubara sum formula for the Casimir pressure between magnetic
metallic mirrors''}

\author{
G.~L.~Klimchitskaya}

\author{
V.~M.~Mostepanenko}
\affiliation{Central Astronomical Observatory at Pulkovo of the Russian Academy of Sciences,
Saint Petersburg,
196140, Russia}
\affiliation{Institute of Physics, Nanotechnology and
Telecommunications, Peter the Great Saint Petersburg
Polytechnic University, St.Petersburg, 195251, Russia}

\begin{abstract}
Recently Gu\'{e}rout {\it et al.} [Phys. Rev. E {\bf 93}, 022108 (2016)] advocated that
the lossless plasma model has to be redefined as the limit of the Drude model when the
relaxation parameter goes to zero.  It was claimed that the previously used plasma model
cannot correctly describe the Casimir pressure between two plates made of both nonmagnetic
and magnetic metals and has  to be replaced with the redefined one. We show that the
suggested redefinition does not satisfy necessary physical requirements imposed on the
dielectric permittivity. We also present a plausible explanation to the fact that the
lossless plasma model describes the Casimir pressure correctly even though it does not
match the optical and electrical properties of metals.
\end{abstract}
\maketitle

It is common knowledge that the experimental data of all precise experiments on measuring
the Casimir force between both nonmagnetic and magnetic metallic surfaces exclude theoretical
predictions of the Lifshitz theory if the low-frequency dielectric permittivity is described
by the Drude model \cite{1,2,3,4,5,6}. The same data are found in good agreement with theory
if the dielectric permittivity at low frequencies is described by the lossless plasma model
\cite{1,2,3,4,5,6}. In both cases the contribution of bound (core) electrons to the
dielectric permittivity $\varepsilon(\omega)$ is found using the tabulated optical data for
the complex index of refraction \cite{7}. Keeping in mind that the presence of ohmic losses in
the dielectric response of metals to real electromagnetic fields of low frequencies is a well
confirmed fact, the exclusion of the Drude model is often considered as puzzling.

Reference \cite{8} discusses this puzzle and arrives at a conclusion that the lossless
plasma model which is commonly used has to be redefined as the limit of the Drude model
\begin{equation}
\chi_{\gamma}(\omega)\equiv\varepsilon_{\gamma}(\omega)-1=
-\frac{\omega_p^2}{\omega(\omega+i\gamma)}
\label{eq1}
\end{equation}
\noindent
when the relaxation parameter $\gamma$ goes to zero ($\omega_p$ is the plasma frequency of
the metal). According to Ref.~\cite{8}, the resulting susceptibility of the plasma model is
\begin{equation}
\chi_{\eta}(\omega)\equiv\lim_{\gamma\to 0}\chi_{\gamma}(\omega)=
-\frac{\omega_p^2}{\omega^2}-i\pi\omega_p^2\delta^{\prime}(\omega),
\label{eq2}
\end{equation}
\noindent
where $\delta^{\prime}(\omega)$ is a derivative of the Dirac $\delta$-function.
This is different from the commonly
used susceptibility of the plasma model \cite{9}
\begin{equation}
\chi_{0}(\omega)=
-\frac{\omega_p^2}{\omega^2},
\label{eq3}
\end{equation}
\noindent
which is obtained from the Drude model (\ref{eq1}) by putting $\gamma=0$ from the outset.
In other words, the Drude model (\ref{eq1}) is a discontinuous function of $\gamma$ at
the point $\gamma=0$, and Ref.~\cite{8} favors the definition of the plasma model by
Eq.~(\ref{eq2}) over that of Eq.~(\ref{eq3}). As stated in Ref.~\cite{8}, ``The lossless
plasma model, with susceptibility $\chi_{0}(\omega)$, does not match the optical and
electrical properties of gold, and it cannot describe correctly the Casimir pressure
between two metallic plates."
According to Ref.~\cite{8}, ``the plasma model can only be considered as an effective model
at high frequencies
$\omega\gg\gamma$. Considered in this manner, it has to be defined as the limit of the
Drude model when $\gamma\to 0$", i.e., by Eq.~(\ref{eq2}).
It should be noted, however, that the redefined plasma
model (\ref{eq2}) introduced in Ref.~\cite{8} also does not match the optical and electrical
properties of gold.
As recognized in Ref.~\cite{8}, the use of the redefined model
(\ref{eq2}) does not solve the Casimir puzzle. Reference \cite{8} considers that an
advantage of their approach is the absence of a discontinuity between the Casimir forces
calculated using the Drude model (\ref{eq1}) and the redefined plasma model (\ref{eq2}).

Below we demonstrate that the suggested susceptibility (\ref{eq2}) does not satisfy some
necessary physical requirements obeyed by the standard dielectric susceptibility (\ref{eq3}).
We also confirm that the lossless plasma model (\ref{eq3}) combined with the contribution of
core electrons does describe correctly the Casimir pressure between two metallic plates even
though it does not match the optical and electrical properties of gold. A plausible explanation
for this fact is presented.

First and foremost we note that the definition of the plasma model as the limiting case of
the Drude model (\ref{eq1}) was used much earlier in Ref.~\cite{10} with the result
\begin{equation}
\chi_{\eta}(\omega)\equiv\lim_{\gamma\to 0}\chi_{\gamma}(\omega)=
-\frac{\omega_p^2}{\omega^2}+i\frac{\omega_p^2}{\omega}\pi\delta(\omega).
\label{eq4}
\end{equation}
\noindent
This made it possible to formally bring the plasma model in agreement with the standard
Kramers-Kronig relation \cite{11}
\begin{equation}
\chi_{\eta}(i\xi)=\frac{1}{\pi}\int_{-\infty}^{\infty}
\frac{\omega\,{\rm Im}\chi_{\eta}(\omega)}{\omega^2+\xi^2}d\omega.
\label{eq5}
\end{equation}
\noindent
The relation (\ref{eq5}), however, is derived for the functions $\chi_{\eta}(\omega)$
analytic in the upper half-plane of complex $\omega$ and regular at $\omega=0$ or having
a pole of no higher than of the first order. Thus, Eq.~(\ref{eq5}) is not applicable
to the plasma model (\ref{eq3}). The generalized Kramers-Kronig relations valid for the plasma-like
susceptibilities of the form \cite{12}
\begin{equation}
\chi(\omega)=
-\frac{\omega_p^2}{\omega^2}+\sum_{j=1}^{K}
\frac{f_j}{\omega_j^2-\omega^2-ig_j\omega},
\label{eq6}
\end{equation}
\noindent
where $\omega_j\neq 0$ are the resonance frequencies of $K$ core electrons,
$g_j$ are their relaxation frequencies, and $f_j$ are their oscillator strengths
are derived in Ref.~\cite{13}. Here, the parameters of oscillators $\omega_j,\,g_j$,
and $f_j$ are determined from the tabulated optical data for Au, Ni or any other metal.
The permittivity (\ref{eq6}) matches the optical data at $\omega> 2$\,eV but, as well as
the permittivity $\chi_0(\omega)$, does not match the optical data at lower $\omega$.
The generalized Kramers-Kronig relations are given by
\begin{eqnarray}
&&
{\rm Re}\chi(\omega)=\frac{1}{\pi}{\rm P}\int_{-\infty}^{\infty}
\frac{{\rm Im}\chi(\xi)}{\xi-\omega}d\xi-
\frac{\omega_p^2}{\omega^2},
\nonumber \\
&&
{\rm Im}\chi(\omega)=-\frac{1}{\pi}{\rm P}\int_{-\infty}^{\infty}
\frac{{\rm Re}\chi(\xi)+1+\frac{\omega_p^2}{\xi^2}}{\xi-\omega}d\xi,
\nonumber \\
&&
\chi(i\xi)=\frac{1}{\pi}\int_{-\infty}^{\infty}
\frac{\omega\,{\rm Im}\chi(\omega)}{\omega^2+\xi^2}d\omega+
\frac{\omega_p^2}{\xi^2}.
\label{eq7}
\end{eqnarray}
\noindent
Both the susceptibilities (\ref{eq3}) and (\ref{eq6}) satisfy the
generalized Kramers-Kronig
relations (\ref{eq7}) directly with no modification. Using the definition of the
derivative of $\delta$-function in the framework of the theory of distributions
\cite{14}, it is easily seen that
\begin{equation}
\omega\delta^{\prime}(\omega)+\delta(\omega)=0,
\label{eq8}
\end{equation}
\noindent
i.e., Eq.~(\ref{eq2}) used in Ref.~\cite{8} is equivalent to Eq.~(\ref{eq4}) used
in Ref.~\cite{10} (see also Refs.~\cite{13,15}).

In fact the functions (\ref{eq2}) and (\ref{eq4}) cannot be continued to the
upper half-plane
of complex $\omega$ in a consistent way because they contain either the delta
function or its derivative. This is also the reason why the imaginary parts of
the functions (\ref{eq2}) and (\ref{eq4}) cannot be obtained from their real parts
by means of some  dispersion relation. Thus, these functions do not satisfy
the necessary physical conditions required from the dielectric susceptibility \cite{11}.
Eventually, the use of such functions as susceptibilities would result in a violation
of the causality principle.
This invalidates the main result of Ref.~\cite{8} on a derivation of the
Lifshitz-Matsubara sum formula for the Casimir pressure between magnetic metallic
mirrors using the dielectric susceptibility (\ref{eq2}).

Now we comment on the statement that ``The lossless plasma model, with susceptibility
$\chi_{0}(\omega)$,\,\ldots cannot describe correctly the Casimir pressure between metallic
plates." However, currently the situation is that the measurement data of all precise experiments
with both nonmagnetic and magnetic metals \cite{1,2,3,4,5,6} are in agreement with theoretical
predictions of the Lifshitz theory at more than 90\% confidence level \cite{15} if the
conduction electrons are described by the plasma model (\ref{eq3}) and core (bound)
electrons by the optical data \cite{7}. Effectively this means that the plasma-like
dielectric permittivity (\ref{eq6}) is used. In experiments of Refs.~\cite{1,2,3,4,5,6}
the difference in theoretical predictions which include and neglect the relaxation properties
of free electrons is of only a few percent. Recently, however, one more, differential, force
measurement has been performed between patterned magnetic metals \cite{16,17} based on
the novel idea proposed in
Refs.~\cite{18,19,20}. In the experiment of Refs.~\cite{16,17} the theoretical predictions
using the Drude model (\ref{eq1}) and the plasma model (\ref{eq3}) at low frequencies differ
by the factor of many hundred. As a result, the theoretical prediction using the plasma
model was confirmed, and the use of the Drude model was excluded conclusively.

The question arises on how this fact can be combined with the correct statement of Ref.~\cite{8}
that ``The lossless plasma model, with susceptibility
$\chi_{0}(\omega)$, does not match the optical and electrical properties of gold,\,\ldots ",
as opposed to ``much better motivated lossy Drude model". We agree with the authors of
Ref.~\cite{8} that this question remains to be solved. It is not true, however, that the very good
agreement of the measurement data of many experiments with theoretical predictions using the
plasma model (\ref{eq6}) ``correspond to inconsistent calculations"\,\ldots ``which has to be
corrected accordingly". The point is that the Casimir force originates from fluctuating
electromagnetic field having a zero expectation value. As to the optical and electrical properties,
they are measured as a response of gold, nickel and other metals to real electromagnetic fields
with nonzero expectation values. The classical theory of electromagnetic fluctuations usually
{\it postulates} similar reaction of a physical system to real and fluctuating fields \cite{11}.
Recent developments in theoretical and experimental investigation of the Casimir effect discussed
above suggest that this might be not the case for some types of quantum fluctuations.
This assumption is supported by the fact that the Lifshitz theory using the Drude model violates
the Nernst heat theorem for both nonmagnetic and magnetic metallic test bodies with perfect crystal
lattices \cite{21,22,23}. The redefined plasma model (\ref{eq2}) and (\ref{eq4}) also violates
the Nernst heat theorem which is satisfied by the commonly used plasma model (\ref{eq3}) and the
generalized plasma model (\ref{eq6}).
Only future investigations will show
whether the Casimir puzzle can be explained along these lines.

To conclude, the redefined plasma model (\ref{eq2}) introduced in Ref.~\cite{8} is not only in
disagreement with the optical and electrical properties of metals at low frequencies, similar
to the plasma model (\ref{eq3}), but it results in violation of thermodynamics and
does not possess necessary physical properties required from dielectric susceptibilities, and,
specifically, does not satisfy the Kramers-Kronig relations. Thus, it cannot be considered
as an alternative to the commonly used plasma model (\ref{eq3}).

\end{document}